\documentclass[10pt]{article}
\usepackage{amssymb}
\usepackage{graphicx}
\graphicspath{{figure/}} %find in the folder figure the figures

\usepackage[mathlines]{lineno}% Enable numbering of text and display math
\modulolinenumbers[5]% Line numbers with a gap of 5 lines
\linenumbers\relax % Commence numbering lines

\usepackage[utf8]{inputenc}
\usepackage[francais, english]{babel} 
\usepackage[T1]{fontenc}
\usepackage{lmodern}
\usepackage{amsmath}
\usepackage{caption}

\usepackage{plain} %empty headpage and pages at bottom center
\usepackage{subcaption}
\usepackage{multirow}
\usepackage{gensymb}

\usepackage{tabularx}

\usepackage[compact]{titlesec}
\titlespacing*{\section}{0pt}{0pt plus 3pt}{3pt}
\titlespacing*{\subsection}
{0pt}{5.5ex plus 1ex minus .2ex}{4.3ex plus .2ex}

\usepackage[a4paper]{geometry}%  page type
\usepackage[cm]{fullpage} % use full page
\usepackage{float} % Required for tables and figures in the multi-column environment - they need to be placed in specific locations with the [H] (e.g. \begin{table}[H])
\usepackage[labelfont=it]{caption}
\usepackage{array}

\title{\fontsize{20pt}{10pt}\selectfont\textbf{On EM Reconstruction of a Multi Channel Shielded Applicator for Cervical Cancer Brachytherapy: A Feasibility Study. }}
\author{ Tho,Daline \up{1,2},Emmanuel Racine \up{1,2}, Harry Easton \up3, William Y. Song \up4 and Luc Beaulieu \up{1,2}\\
\normalsize\up1 Département de radio-oncologie et Centre de recherche du CHU de Québec, CHU de Québec,Canada\\
\normalsize\up2 Département de physique, de génie physique et d'optique, et Centre de recherche sur le cancer, Université Laval, Canada\\
\normalsize\up3 Departement of Radiation Oncology, Sunnybrook Health Sciences Centre Toronto, Canada\\
\normalsize\up4 Department of Radiation Oncology, Virginia Commonwealth University, Richmond, VA}
\begin{document}
\maketitle
Corresponding author: Luc Beaulieu\\
email: Luc.Beaulieu@phy.ulaval.ca 
%===========================================

%===============resume=======================
\begin{abstract}
Electromagnetic tracking (EMT) is a promising technology for automated catheter and applicator reconstructions in brachytherapy. In this work, a proof-of-concept is presented for reconstruction of the individual channels of a shielded tandem applicator dedicated to intensity modulated brachytherapy. All six channels of a straight prototype was reconstructed and the distance between two opposite channels was measured. A study was also conducted on the influence of the shield  on the data fluctuation of the EMT system. The differences with the CAD specified dimensions  are under 2 mm. The pair of  channels which has one of it more distant from the generator have higher  inter-channel distance with higher variability. In the first 110 cm reconstruction, all inter-channel distances are within the geometrical tolerances.  According to a paired Student t-test, the data given by the EM system with and without the shield applicator tip are not significantly different. This study shows that the reconstruction of channel path within the mechanical accuracy of the applicator is possible.
 \end{abstract}
%================INTRODUCTION===========================
%\begin{multicols}{2}
\section {Introduction}
The emergence of electromagnetic tracking (EMT) technologies is currently paving the way for a variety of novel applications in the medical field. It is used in surgical  procedures \cite{2006SPIE.6141..152N} and gives accurate  measurements \cite{Lugez2015} \cite{NDI}. The Aurora system is composed of a field generator which produces a known EM field within a known volume inside of which, a small sensor is placed. The system uses mutual induction to compute the position of the sensor. Brachytherapy is used to treat numerous body sites including prostate, breast  and gynecology. Many different applications of minimally invasive image-guided brachytherapy has been demonstrated like magnetic resonance tomography, computed tomography or ultrasound imaging data. All of those reconstruction imaging method have their limitations in term of accuracy \cite{Richart:2015aa} \cite{Smith:aa} \cite{Limbacher:aa}.

It is possible to use the EMT to perform reconstructions \cite{DAMATO:2014}. At the same time, shielded applicators for brachytherapy treatment are making their way to improve dosimetric results \cite{Webster2013}. According to the literature, EMT tracking accuracy is sensible to conductive metal which causes distorsion in the magnetic field \cite{BOUTALEB:2014}\cite{Zhou:2013}. 

In this technical note, a proof-of-concept is presented for reconstruction of the individual channels of a prototype shielded tandem applicator. The technique has the potential to considerably speed up planning and quality assessment tasks in the clinic, particularly in this context when imaging of such applicator might be impossible due to artifacts.  The purpose of this work is to show the feasibility of EMT for automated multi-channel shielded applicator reconstruction. 

\section{Materials and Methods}%=================methode
\subsection{Straight shielded applicator}
The applicator used in this work corresponds to the straight, intra-uterine portion of a shielded multi-channel tandem prototype previously described by Han \textit{et al}. (Figure 1a) \cite{Han2014666}. All 6 channels were reconstructed (Figure 1b). The inter-channel distances of each opposite channel pair, measured along the complete 14 cm length, was used as a metric for validation of EMT reconstruction i.e. absence of geometric distortion. Figure 1c shows that this nominal distance is 4 mm. The shielded part is composed of a MR-compatible non-ferromagnetic alloy made of 95$\%$tungsten, 3.5$\%$ nickel, and 1.5$\%$ copper) as described in Han \textit{et al.} \cite{Han2014666}. 
%-------------------------fig 1 -------------------------------------------

\begin{figure} [H]%working if you put H
\centering

       \includegraphics[width=13cm]{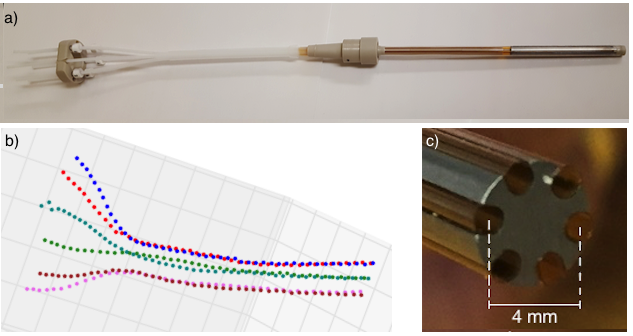}\label{demo2}

\caption{Straight shielded prototype with its 6 channels reconstruction. a) The prototype , b) the EMT reconstruction data and c) the applicator tip illustrating all 6 channels.} 
\end{figure}

\subsection{EM tracking system}
\begin{figure} [H]%working if you put H
	\centering
 \includegraphics[width=13cm]{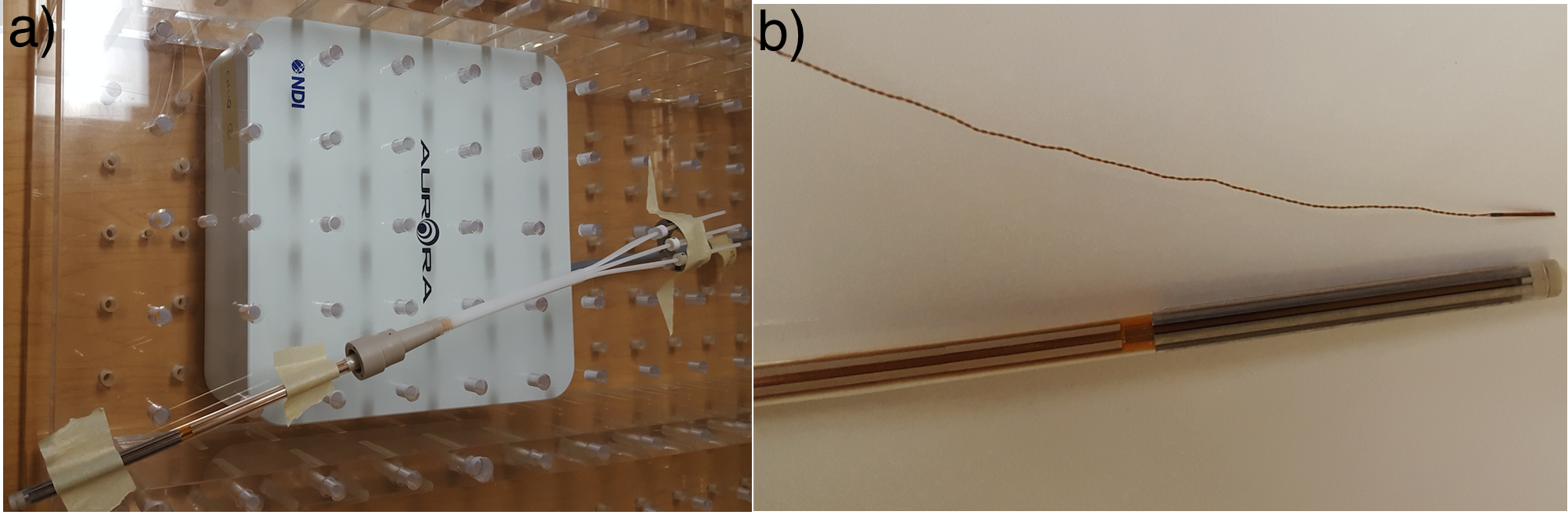}\
\caption{Applicator and the sensor. a) a minimal perturbation set-up was in place using only wood and plastic.b) a close up of the measurement configuration with the shielded applicator place in proximity of the sensor.} \label{applicator_sensor}
\end{figure}

The shielded tandem applicator was reconstructed using the Aurora® V3 EMT system (NDI, Waterloo, Ontario, Canada) using model 610090 5 degree-of-freedom (DOF) sensor (Figure \ref{applicator_sensor}). The sensor is a cylinder of 0.8 mm diameter and 11 mm long (Figure \ref{applicator_sensor}). 

%-------------fin demo 2--------------------------------------------------------
\subsection{Effect of shield on data fluctuation}
Repeated measurements of EMT sensor positions within the detection volume, with and without the shielded applicator present, was used to study the potential perturbation effect of the shield on the sensor reading (Figure \ref{applicator_sensor}). The effective detection volume is a cube of 50 cm side \cite{NDI}. Nine different positions (one near each corner of the sensible cube volume and the center) were taken each time with the sensor directly on the applicator and without the applicator. Each measurement last 10 s with a frequency of 40 measurements/s. On Figure \ref{applicator_sensor}, the shielded end of the applicator is place next to the sensor.

\section{Results and discussion} \vspace{-0.1cm}%===================================RESULTAT 
\subsection{Straight shielded applicator}
The distance between  the center of 2 opposite  channels is nominally 4 mm from center-to-center (Figure 1c). Table \ref{difference-center-to-center} reports the experimental distances computed from the EMT reconstructed channel. The average distance from each pair met the nominal distance within one standard deviation. It is also important to note that each hole is 1.4 mm in depth (or $\pm$ 0.7 mm from its center).  

\begin{table}[H]
\caption{Differences from center to center opposite pair of channel.}\label{difference-center-to-center}
\centering
\begin{tabular}{cccccccc}

Pair&Average&Standard deviation\\
number&(mm)&(mm)\\
\hline
3-6&4.33&0.40\\
2-5&3.88&0.26\\
1-4&4.14&0.35\\
\hline
\end{tabular}
\end{table}
The distance were also computed along the reconstruction path.

\begin{figure} [H]%working if you put H
\centering
     \includegraphics[width=9cm]{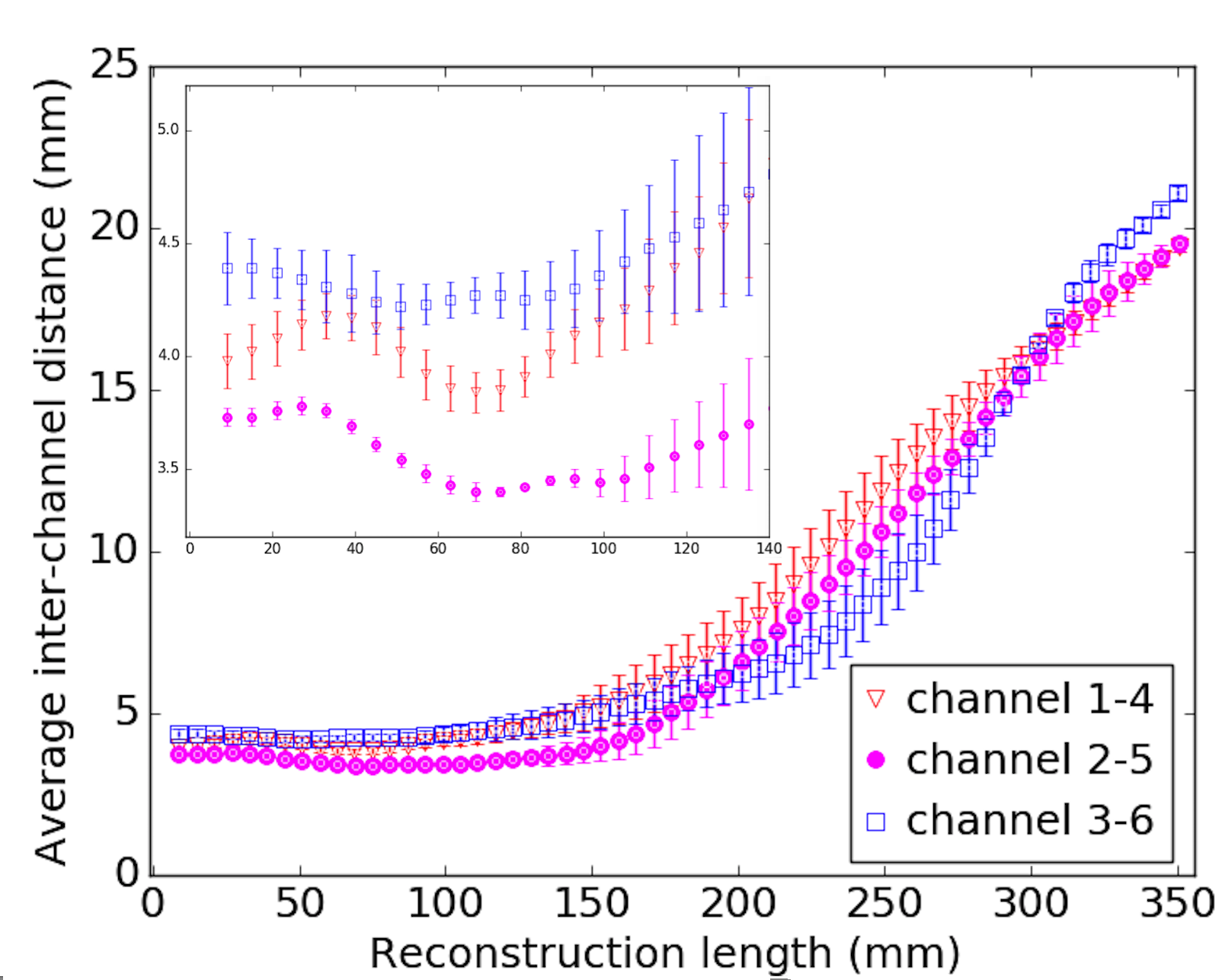}
\caption{Computed inter-channel distance along the reconstruction path. The insert correspond to the shielded-only portion. The dashed line are the geometrical error tolerance. The nominal inter-channel distance is 4 mm.}\label{interchanneldistance}
\end{figure}

On Figure \ref{interchanneldistance}, one can see that most of the shielded part has its inter-channel distance within the geometrical tolerance of the setup.  The channel  which were on the side during the reconstruction have their  inter-channel distance  always below the nominal value. Meaning that during the reconstruction, the sensor tend to be on the  side of the channel nearer to the middle of the applicator.

\subsection{Stability of the reading with a shield present}
With the nine different positions, a paired t-test was made. According to it, the data given by the EMT system with and without the shielded applicator tip are not significantly different (Table \ref{tipeffect}). A p-value < 0.05 was considered statistically significant. The absolute maximum difference  is 0.05 mm which was for the Z axis. The standard deviation of all the differences between the acquired  position with and without the shield near are sub-millimetric which is comparable with the standard deviation of the computed position.

\begin{table}[H]
\caption{Paired t-test p-value to study the effect of the applicator on the EM tracking system.}\label{tipeffect}
\centering
\begin{tabular}{ccccccc}
Position&p&|$\Delta| $(mm)&$\sigma$ (mm)\\
\hline
X&0.09&0.02&0.04\\
Y&0.72&0.01&0.05\\
Z&0.38&0.05&0.03\\
\hline
\end{tabular}
\end{table}

\subsection{Limitations}
The off-the-shelf sensor used in this work is smaller than the channel diameter of the applicator. This could explain part of the fluctuation seen in the interchannel distance measurements. Furthermore, the applicator geometrical tolerance, corresponding to the channels radius, has to be taken into account. As such, a maximum of $\pm$ 0.7 mm difference has been considered in the analysis, represented by the dashed lines in Figure \ref{interchanneldistance}. After the first 110 cm of reconstruction, the measured interchannel-distances increases. As we approach the end of the shielded portion of the applicator, channels start to move away from each other and the sensor with it. Also, since the sensor is smaller than the channel radius, the effect of gravity on the sensor position within the channels cannot be ruled out.

While this is a shielded applicator, each channel is not completely closed (Figure 1c) and material is non-ferromagnetic. As such the results obtained here might not be generalizable to all shielded geometries, in particular those that would fully enclose the sensor, acting as a Faraday cage.

 \section{Conclusion}
It appears feasible to use EMT technology to reconstruct shielded applicators, such as the one presented here. The observed deviations shows no statistically significant effect on the presence of the shielding material in the detection volume (static configuration). This study demonstrates that it is possible to reconstruct the channel path within the mechanical accuracy of the applicators. 
\section*{Acknowledgments}
This work was supported by the National Sciences and Engineering Research Council of Canada (NSERC) via the NSERC-Elekta Industrial Research Chair. Daline Tho acknowledges support from the Medical Physics Training Network CREATE NSERC grant \# 432290.

 %===========================================   
%REFERENCES ============================================  

\bibliographystyle{ama}

%\end{multicols}

\end{document}